\documentclass[10pt,twocolumn]{article}
\usepackage{epsbox}
\usepackage{times}
\usepackage{enumerate}
\usepackage{amsmath,amssymb,amsfonts}
\usepackage{graphicx}
\usepackage{algorithm}
\usepackage{algpseudocode}
\usepackage{enumerate}
\usepackage{textcomp}
\usepackage{epstopdf} 
\usepackage{subfloat}
\usepackage[caption=false]{subfig}
\usepackage{pgfplots}
\pgfplotsset{compat=newest,compat/show suggested version=false}
\usepackage{hyperref}
\usepackage{algpseudocode}
\usepackage{url}
\usepackage{ifpdf}
\usepackage{array}
\usepackage{bm}
\usepackage[justification=centering]{caption}
\usepackage{caption}
\usepackage{tikz,pgfplots,pgfplotstable}
\usetikzlibrary{pgfplots.groupplots}
\usetikzlibrary{fit,arrows, backgrounds, calc, hobby, positioning}
\usetikzlibrary{shapes.geometric}
\usepackage{float}
\usepackage{paralist}
\newcommand{\tacc}{\textit{TACC\_Stats}}%{{\tt TACC\_Stats}}
\newcommand{\sysname}{\textit{dynamicMF}}

\newcommand{\norm}[1]{\left\lVert#1\right\rVert}

\usepackage{booktabs} % For formal tables

\begin{document}
\title{dynamicMF: A Matrix Factorization Approach to Monitor Resource Usage in High Performance Computing Systems}

\author{
	Niyazi Sorkunlu\\
	Computer Science and Engineering \\
		University at Buffalo, \\
		State University of New York\\
		Buffalo, New York 14260\\
		\texttt{niyaziso@buffalo.edu}\\
	\and
	Duc Thanh Anh Luong\\
	Computer Science and Engineering \\
		University at Buffalo, \\
		State University of New York\\
		Buffalo, New York 14260\\
		\texttt{ducthanh@buffalo.edu}
	\and
	Varun Chandola\\
	Computer Science and Engineering \\
		University at Buffalo, \\
		State University of New York\\
		Buffalo, New York 14260\\
		\texttt{chandola@buffalo.edu}
	}

\maketitle

\begin{abstract}

High performance computing (HPC) facilities consist of a large number of interconnected computing units (or nodes) that execute highly complex scientific simulations to support scientific research. Monitoring such facilities, in real-time, is essential to ensure that the system operates at peak efficiency. Such systems are typically monitored using a variety of measurement and log data which capture the state of the various components within the system at regular intervals of time.
As modern HPC systems grow in capacity and complexity, the data produced by current resource monitoring tools is at a scale that it is no longer feasible to be visually monitored by analysts.
We propose a method that transforms the multi-dimensional output of resource monitoring tools to a low dimensional representation that facilitates the understanding of the behavior of a High Performance Computing (HPC) system. 
The proposed method automatically extracts the low-dimensional signal in the data which can be used to track the system efficiency and identify performance anomalies. The method models the resource usage data as a three dimensional tensor (capturing resource usage of all compute nodes for difference resources over time). A dynamic matrix factorization algorithm, called {\em dynamicMF}, is proposed to extract a low-dimensional temporal signal for each node, which is subsequently fed into an anomaly detector. Results on resource usage data collected from the {\em Lonestar 4} system at the Texas Advanced Computing Center show that the identified anomalies are correlated with actual anomalous events reported in the system log messages.
\end{abstract}

%\keywords{HPC \and  anomaly detection \and  system monitoring \and  performance profiling \and  matrix factorization}

\maketitle
\section{Introduction}
High performance computing (HPC) is at the forefront of scientific discovery and engineering innovation. To support the growing computing demands in these areas, modern HPC systems have seen a rapid transformation in terms of compute capacity and the underlying complexity. To ensure that an HPC system is operating at its peak efficiency, analysts typically rely on system monitoring tools to monitor the system performance and identify anomalies. 

Several tools exist for collecting and visualizing resource usage data from large scale HPC installations (e.g. Texas Advanced Computing Center \tacc ~\cite{Evans2014}, XSEDE Metrics on Demand or XDMoD~\cite{Palmer2015}, etc.). Such tools can produce large amounts of high dimensional resource usage data at a high temporal frequency for each computational node in the system. The data collected by such tools provides a real-time view of the performance of the system which is typically fed into a visual interface (e.g., XDMoD~\cite{Palmer2015}, OVIS~\cite{Brandt:2009}, Bright Cluster~\cite{BrightComputing}, etc.). However, such data is typically large and unwieldy, and visual monitoring is often challenging and inefficient.% and can be analyzed to identify system performance anomalies.

Existing automated methods typically monitor each system resource or state for every compute node independently for potential deviations or anomalies~\cite{Peiris:2014}, relying on a pre-defined threshold. Such methods can identify only those anomalous scenarios in which an individual node exhibits significant deviation for individual resources. These methods often miss anomalies that are under the threshold. Such anomalies, are weakly manifested across several nodes and multiple system resources and can be potentially detected by understanding the interactions between the different aspects of the system. A tensor decomposition based method~\cite{Sorkunlu:2017} jointly models the interactions across the nodes, resources, and time, and produces a single time varying signal that can be used for tracking the overall system performance. However, the output does not allow for finer resolution analysis, for instance, identifying the specific nodes that are performing anomalously.

\begin{figure}[htbp]
	\centering
	\captionsetup{justification=raggedright,singlelinecheck=false}
	\includegraphics[width=0.48\textwidth]{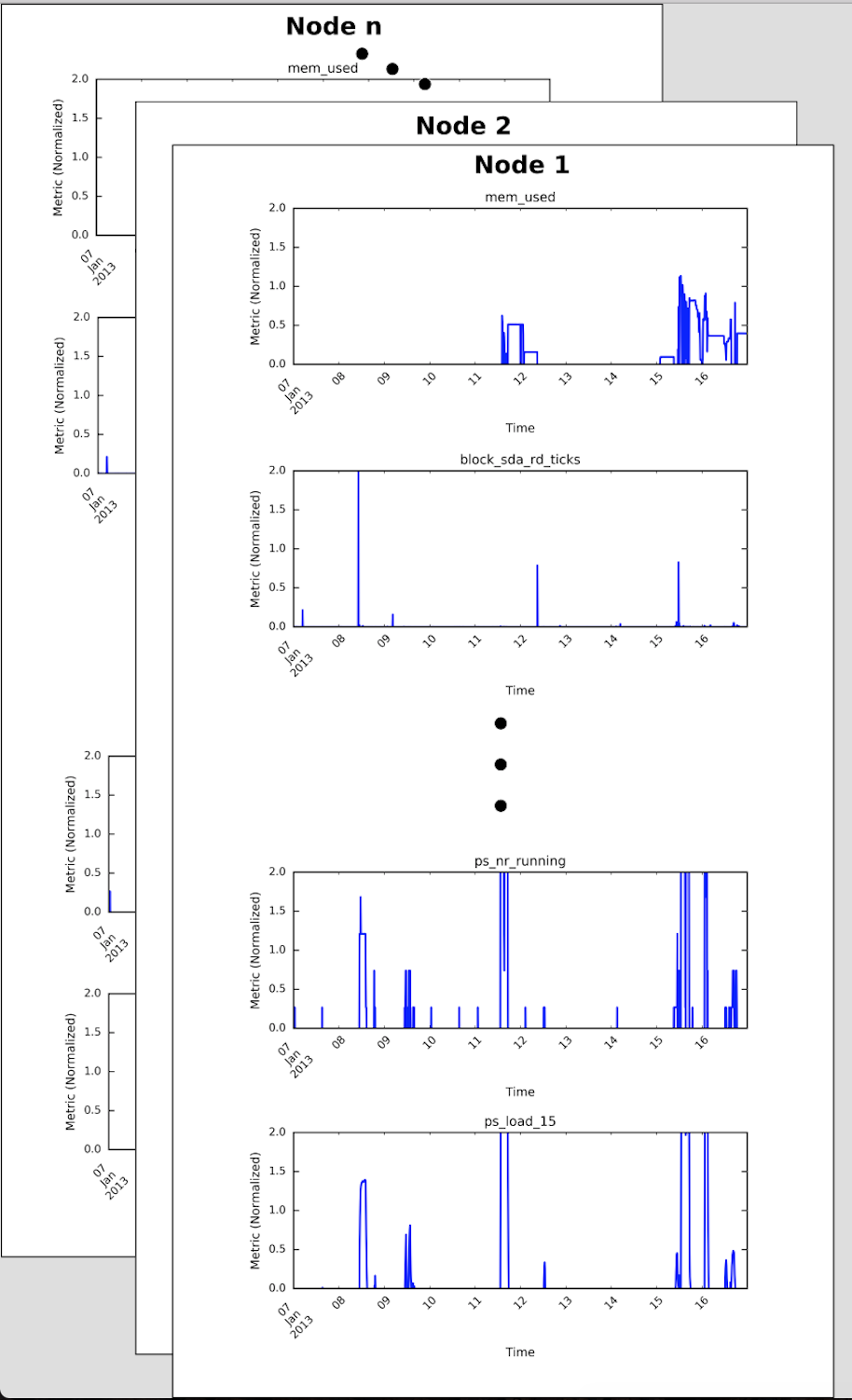}
	\caption{Typical resource usage data collected by tools such as \tacc. Each panel corresponds to a compute node and consists of time-varying usage metrics for a variety of system resources.}
	\label{fig:overview}
\end{figure}

%Resource usage data is typically large and unwieldy. For instance, in this paper we analyze resource usage data from the {\em Lonestar4} cluster at the {\em Texas Advanced Computing Center}.
%
In this paper, we propose a method for understanding the system behavior at the node resolution. The key assumption here is that the observed system behavior, captured as resource usage information by tools such as \tacc, can be decoupled into node and metric specific behaviors. Further, the node behavior can be decoupled into time-invariant (or static) and time-dependent (or dynamic) terms. This ``decoupling'' is achieved using a dynamic matrix factorization method which operates on the sequence of {\em node}-{\em usage} matrices, collected over time (See Figure~\ref{fig:overview}). The output of the proposed algorithm is a set of low-dimensional representations for the metrics and the nodes that facilitate understanding of the system in multiple ways. In particular, the dynamic node representation allows for tracking the performance of each node. Additionally, we use the proposed algorithm to produce a node level anomaly statistic and show that many performance anomalies, identified from message logs, co-occur with the identified anomalies. 
%With the increasing use of high performance computing (HPC) system in research and analysis, the architecture and computing power of HPC systems are evolving into complex structures which require efficient tools to monitor and quickly identify failures in the systems. 
%Currently, there are various tools for monitoring resource usage of each node in HPC systems such as Texas Advanced Computing Center \tacc~\cite{Evans2014}, XSEDE Metrics on Demand (XDMoD)~\cite{Palmer2015}, etc. 
%These tools can provide many resource usage statistics with high temporal frequency for each computational node. 
%In addition, as the added overhead for producing these statistics is relatively cheap, the data collected by such tools become abundance and they are potential for retrospective analysis to have better understanding of system performance and identify inefficiency in system resource usage.
% 
%HPC monitoring program typically produces resource usage statistics as a three-way tensor data including node, metric (resource usage statistics) and time. This data is illustrated in Figure~\ref{fig:overview}.

The paper is organized as follows. The key problem addressed in this paper is outlined in Section~\ref{sec:algorithm}, along with the proposed model, \sysname. Results on a week long \tacc\ data for the Lonestar 4 system in the Texas Advanced Computing Center are discussed in Section~\ref{sec:experiment}. Section~\ref{sec:discussion} provides a brief overview of related works in this area and conclusion is presented in Section~\ref{sec:conclusion}.
\label{sec:introduction}

\section{Proposed Methodology}
\label{sec:algorithm}
We describe the proposed \sysname\ algorithm in this section. The algorithm is motivated by the fact that, while the performance of an HPC system is captured for many nodes and in the context of several resources, the metrics and the nodes exhibit a clustering pattern: several resources produce correlated usage metrics and many nodes behave similarly, either due to their computing specifications or the workload.
\subsection{Problem Setting}
We consider an HPC system consisting of $N$ compute nodes. A resource usage monitoring program, such as \tacc\, periodically reports usage data corresponding to $M$ different resource usage metrics (CPU consumption, memory usage, network usage are three typical examples). At a given time $t$, the variable $z_{nmt}$ denotes the $m^{th}$ resource's usage for node $n$ in the time interval ending at time $t$. We collect all of these variables at time $t$ in a matrix ${\bf Z}_t \in \mathbb{R}^{N \times M}$, such that $Z_t[n,m] = z_{nmt}$.
\subsection{Proposed Model}
The proposed model assumes that the nodes and the metrics can be represented as {\em vectors} in two distinct (and unobserved/latent) $K$-dimensional spaces ($K \ll N,M$), respectively. Thus each metric is represented as a $K$-dimensional vector, ${\bf v}_m \in \mathbb{R}^K$, where each of the $K$ dimensions denote a canonical characteristic of the corresponding metric. For the nodes, we assume two representations. The first representation is static over time, denoted as ${\bf \bar{u}}_n \in \mathbb{R}^K$, and characterizes the base behavior of the node. The second representation is time-dependent, denoted as ${\bf \widehat{u}}_{nt} \in \mathbb{R}^K$, and captures the dynamic behavior of the node. 

Assuming that the above representations, including $\{{\bf v}_m\}_{m=1}^M$, $\{{\bf \bar{u}}_n\}_{n=1}^N$, $\{{\bf \widehat{u}}_{nt}\}_{n=1,t=1}^{N,T}$, are known, the observed behavior for each node and metric combination, i.e., $z_{nmt}$, is generated by a functional interaction between the three representations. In this paper, we assume a linear interaction of the following form:
\begin{equation}
  z_{nmt} = ({\bf \bar{u}}_n\odot\widehat{\bf u}_{nt})^\top{\bf v}_m
  \label{eqn:interactionsingle}
\end{equation}
where $\odot$ is an element-wise product. For the entire resource usage matrix, the interaction can be written as:
\begin{equation}
  {\bf Z}_t = ({\bf \bar{U}} \odot {\bf \widehat{U}}_t){\bf V}^\top 
  \label{eqn:interactionfull}
\end{equation}
${\bf V} \in \mathbb{R}^{M \times K}$ consists of the latent representations of the $M$ metrics, i.e., ${\bf V} = [{\bf v}_1, {\bf v}_2, \ldots {\bf v}_M]^\top$. ${\bf \bar{U}} \in \mathbb{R}^{N \times K}$ consists of the static latent representations of the $N$ nodes, i.e, ${\bf \bar{U}} = [\bar{\bf u}_1, \bar{\bf u}_2, \ldots {\bf \bar{u}_N}]^\top$. ${\bf \widehat{U}}_t \in \mathbb{R}^{N \times K}$ consists of the dynamic latent representations of the $N$ nodes at time $t$, i.e, ${\bf \widehat{U}_t} = [\widehat{\bf u}_{1t}, \widehat{\bf u}_{2t}, \ldots {\bf \widehat{u}_{Nt}}]^\top$. Figure~\ref{fig:model} illustrates the interaction between components of the proposed model to generate the observation for each metric and node pair in a given time window.% In this figure, $z_{t,n,m}$ is the 

\subsection{Inference Problem}
\label{subsec:problem}
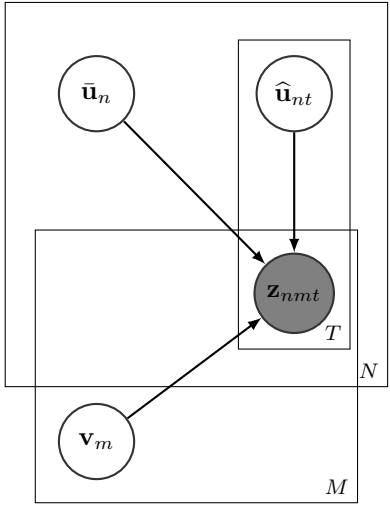
\begin{figure}[htbp]
  \centering
  \begin{tikzpicture}{scale=0.8}
    \tikzstyle{main}=[circle, minimum size = 10mm, thick, draw =black!80, node distance = 16mm]
    \tikzstyle{connect}=[-latex, thick]
    \tikzstyle{box}=[rectangle, draw=black!100]
    \node (z) [main,xshift=10mm,fill=gray]{${\bf z}_{nmt}$};
    \node (uhat) [main,above =of z]{$\widehat{\bf u}_{nt}$};
    \node (ubar) [main,left =of uhat]{$\bar{\bf u}_n$};
    \node (v) [main,below =of ubar,yshift=-20mm]{${\bf v}_m$};

    \path 
    (uhat) edge [connect] (z)
    (ubar) edge [connect] (z)
    (v) edge [connect] (z);
    \node[box, inner sep=3mm,draw=black!100, fit = (z) (v)] (metrics) {};
    \node[box, inner sep=2mm,draw=black!100, fit = (z) (uhat) ]  (dynamicu) {};
    \node[box, inner sep=7mm,draw=black!100, fit = (z) (uhat) (ubar)]  (allu) {};
    \node[anchor=south east] at (metrics.south east) {\footnotesize $M$};
    \node[anchor=south east] at (dynamicu.south east) {\footnotesize $T$};
    \node[anchor=south east] at (allu.south east) {\footnotesize $N$};
  \end{tikzpicture}
  \captionsetup{justification=raggedright,singlelinecheck=false}
  \caption{Illustration of the assumed interactions between the observed data (${\bf z}_{nmt}$) and the latent variables ($\widehat{\bf u}_{nt}, \bar{\bf u}_n, {\bf v}_m$) at a given time $t$ for node $n$ and metric $m$ in the proposed \sysname\ model.}
  \label{fig:model}
\end{figure}

The proposed generative model (See Figure~\ref{fig:model}) shows how the observations are generated through interactions between the three low-dimensional representations. However, given that these representations are unknown, the actual problem here is that of {\em inference}, i.e., given the observed resource usage data, $\{{\bf Z}_t\}_{t=1}^T$, estimate ${\bf V}, \bar{\bf U}$, and $\{\widehat{\bf U}_t\}_{t=1}^T$.

Each of the inferred entities provides a distinct insight into the system performance. The latent metric characteristics (${\bf v}_m$) map the metrics into a common space, where one can understand the similarity between the various metrics as inferred from the data. For instance, one would expect all metrics pertaining to the {\em network} resource appear similar in the latent space. However, how would {\em input/output} related metrics compare with the network metrics? Such relations are quantifiable using the latent space representation. Similar insights can be derived for the nodes using the static node characteristics ($\bar{\bf u}_n$).
However, the dynamic node characteristics ($\widehat{\bf u}_{nt}$) is the key output of the proposed model that allows for tracking the behavior of each node over time using a few latent characteristics. The information in matrices $\{\widehat{\bf U}_t\}_{t=1}^T$ is obtained from the observed resource usage data after ``explaining away'' the static node and metric behavior. Thus, the residual $K$-dimensional signal is expected to better provide information about the true dynamic behavior of each node.

\subsection{Estimating Latent Representations}
We pose the problem of estimating the latent entities, ${\bf V}, \bar{\bf U}$, and $\{\widehat{\bf U}_t\}_{t=1}^T$, as an optimization problem that minimizes the error between the observed data ($\{{\bf Z}_t\}_{t=1}^T$) and reconstruction obtained in~\eqref{eqn:interactionfull}. The optimization problem can be written as:

\begin{eqnarray}
  &&\min \sum_{t=1}^T \norm{{\bf Z}_t - (\bar{\bf U} \odot \widehat{\bf U}_t) {\bf V}^T}_F^2
	\label{eqn:optimization}	
	\\
	&&\text{ with respect to $\bar{\bf U}, {\bf V}, \{\widehat{\bf U}_t\}_{t=1}^T$}
	\nonumber	
\end{eqnarray}
where $\norm{\cdot}_F$ represents the Frobenius norm.

\subsection{Optimization}
\label{subsec:optimization}
In order to solve the optimization problem as stated in~\eqref{eqn:optimization}, we use Adam algorithm~\cite{kingma2014adam} as the main building block to jointly optimize over the entire set of variables. The Adam algorithm performs first-order gradient-based optimization of stochastic objective functions, based on adaptive estimates of lower-order moments, and has been highly successful in the training of deep neural networks. Typically, as a gradient-based method, one step of Adam algorithm only requires one pass over the data. In addition, its optimization with momentum allows the algorithm to work well for mini-batches of data instead of the entire dataset. Beside its fast computation, this algorithm is also efficient in terms of memory usage and well-suited for training model with many parameters. Algorithm~\ref{alg:full_optimization} provides the pseudo-code for our approach.

\begin{algorithm}
\begin{algorithmic}
%\State \textbf{Require:} $\alpha$ (step size)
%\State \textbf{Require:} $\beta_1, \beta_2 \in [0, 1]$ (parameters of exponential moving averages)
\State \textbf{Initialization:} initialize the values of $\bar{\bf U}$, ${\bf V}$ and $\{\widehat{\bf U}\}_{t=1}^T$
\For{$iter \in \{1, \cdots, max\_iter\}$} 
	%\State Optimize $L$ with respect to $\bar{U}, V, \{\widehat{U}_t\}_{t=1}^T$
	\State Update $\bar{\bf U}, {\bf V}, \{\widehat{\bf U}_t\}_{t=1}^T$ simultaneously using one step of Adam optimizer 
\EndFor
\end{algorithmic}
\caption{Adam Optimizer}
\label{alg:full_optimization}
\end{algorithm}

For the experiments conducted in this paper, we implement the \sysname\ algorithm using TensorFlow~\cite{abadi2016tensorflow}. The hyper-parameters of the Adam optimizer are set by following the suggestion in the original paper~\cite{kingma2014adam}, i.e. $\alpha=0.001$ (step size), $\beta_1=0.9$, $\beta_2=0.999$ (decay rate in exponential moving average).

\subsection{Anomaly Detection}
\label{subsec:anomaly}
By using the resulting latent values $\bar{\bf U}$, $\{\hat{\bf U}_t\}_{t=1}^T$ and ${\bf V}$, we can derive an anomaly score for node $n$ at time step $t$ which can be defined as a discrepancy between model and observation, i.e.:
\begin{equation}
{\bf a}_{nt} =\frac{1}{m}\sum_{m}^{M} \left | z_{nmt} - ({\bf \bar{u}}_n\odot\widehat{\bf u}_{nt})^\top{\bf v}_m \right |
\label{eq:residual} 
\end{equation}

%Put Z difference 

%\section{Anomaly Detection}
%\input{section/anomaly.tex}

\section{Experimental Results}
\label{sec:experiment}
In this section, we present the experimental results by applying the \sysname algorithm to analyze real resource usage logs for a large HPC system, and identify potential system anomalies.
%In section~\ref{subsec:data}, we describe the HPC system logs data that we use for our experiments. 
%In section~\ref{subsec:implementation}, we discuss the implementation details of our model. 
%In section~\ref{subsec:result}, we present the experimental results.
%
\subsection{Dataset}
\label{subsec:data}
The dataset used in the experiments is a subset of data obtained from Lonestar 4 cluster located at the Texas Advanced Computing Center (TACC)\footnote{\url{https://portal.tacc.utexas.edu/archives/lonestar4}}. This cluster consists of $1888$ computing nodes from which the records of $1709$ nodes are used in this study. The resource usage data is collected using the \tacc\ system monitor~\cite{Evans2014} which records various resources usage statistics at each computational node every $10$ minutes. In our experiments, we use a set of $86$ resource usage statistics with a resolution of $10$ minutes from 1:10:01 March $1^{st}$ 2013 to 23:40:01 March $7^{th}$ 2013. Totally, we have data of 1000 timesteps that will be used for analysis in our experiments. See table~\ref{tab:metrics} for the complete list of resource usage statistics used in this paper. As different statistics have different scale and units, we have normalized them to have zero mean and unit variance.

\begin{table}[ht]
\centering
	\begin{tabular}{|p{1in}|p{2in}|}
	\hline
	{\bf Component} & {\bf Resource usage metrics}\\
	\hline
	\hline
	CPU & user, nice, system, idle, iowait, irq, softirq\\
	\hline
	I/O&rd\_ios, rd\_merges, rd\_sectors, rd\_ticks, wr\_ios, wr\_merges, wr\_sectors, wr\_ticks\\
	\hline
	Lustre /scratch, /work & read\_bytes, write\_bytes, dirty\_pages\_hits, dirty\_pages\_misses, ioctl, open, close, mmap, seek, fsync, setattr, truncate, getattr, statfs, alloc\_inode, setxattr, getxattr, listxattr, inode\_permission, readdir, lookup, link, unlink, rename\\
	\hline
	Lustre network & tx\_msgs, rx\_msgs, tx\_bytes, rx\_bytes,\\
	\hline
	Virtual Memory & pgpgin, pgpgout, pgalloc\_normal, pgfree, pgactivate, pgdeactivate, pgfault, pgmajfault, pgrefill\_normal, pgsteal\_normal, pgscan\_kswapd, pgscan\_direct, pginodesteal, slabs\_scanned, kswapd\_steal, kswapd\_inodesteal, pageoutrun, allocstall, pgrotated\\
	\hline
\end{tabular}
\caption{List of resource usage metrics used for experiments.}
\label{tab:metrics}
\end{table}

In addition to resource usage statistics, we also have information of computing jobs that performed in Lonestar 4. In particular, for each computing node that the job was assigned to, the system records its start time and end time. There is also the set of system log messages that outputs log traces from programs running on the HPC system, including the messages from the applications running on the nodes.

\subsection{Optimization performance and effect of different K}
\label{subsec:result}
\begin{figure}[htbp]
  \centering
  \captionsetup{justification=raggedright,singlelinecheck=false}
  \includegraphics[width=0.5\textwidth]{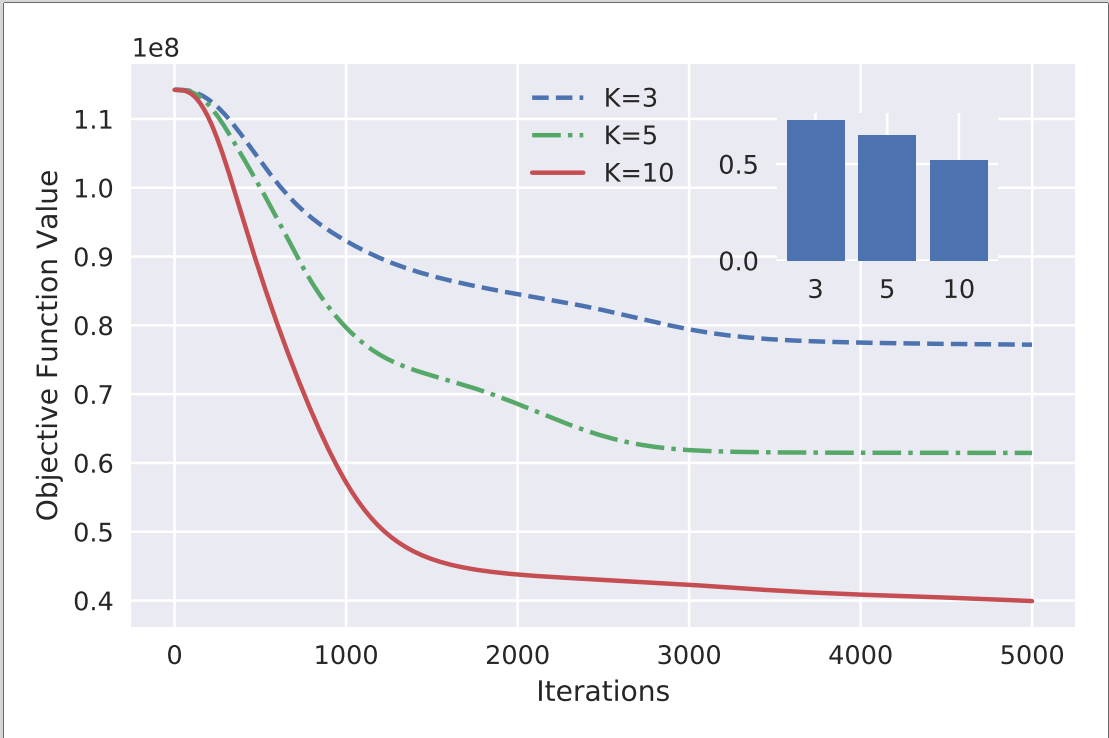}
  \caption{Objective function (See~\eqref{eqn:optimization}) optimization for different number of latent dimensions ($K$). Inset shows the average final objective function value (or error) for $z_{nmt}$. ({\em Best viewed in color.})}
  \label{fig:optimization_error}
\end{figure}

The performance of the optimization algorithm (See Algorithm~\ref{alg:full_optimization}) in estimating the latent representations is shown in Figure~\ref{fig:optimization_error}. Results are shown for $K$ = 3, 5, and 10. For each $K$, the objective function decays with each optimization step and eventually converges.

As expected, using a larger number of latent dimensions results in a better fit. However, larger $K$ also results in a larger number of latent variables to monitor. 
%To balance between the fit and the compactness of the output, we choose $K = 10$ to report the subsequent findings. 
In this paper, we choose $K = 10$ to report the subsequent findings. 
Note that the average error between the model and observations is approximately $0.57$ for $K=10$ (See Figure~\ref{fig:optimization_error} inset).

\subsection{Analyzing latent representations}
\begin{figure}[htbp]
  \centering
  \captionsetup{justification=raggedright,singlelinecheck=false}
  \includegraphics[width=\linewidth]{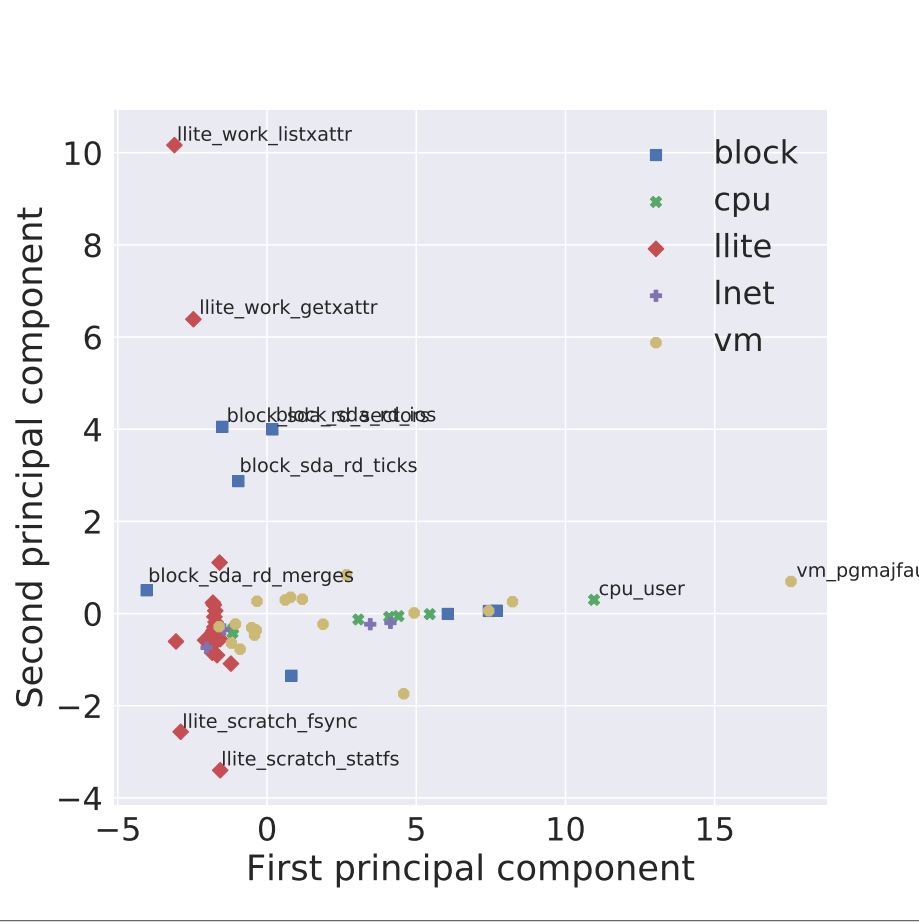}
  \caption{2-dimensional view of the latent representation for metrics estimated by \sysname with $K = 10$. The $K$-dimensional data was reduced to two dimensions using PCA. Metrics belonging to a single coarse grouping ({\tt llite}, {\tt vm}, {\tt cpu}, {\tt cpu}, {\tt lnet}, {\tt block\_sda}) are displayed in the same way. ({\em Best viewed in color.})}
  \label{fig:metric_pca}
\end{figure}

In this section we illustrate how the latent representations obtained using \sysname\ can be used to gain insights about the underlying system. The first latent representation (${\bf V}$) provides a characterization of the metrics. A two-dimensional visualization of the metric representation is shown in Figure~\ref{fig:metric_pca}. We use {\em Principal Component Analysis} (PCA)~\cite{Jolliffe:1986} to reduce the $K$-dimensional data to two dimensions for visualization. 

Figure~\ref{fig:metric_pca} reveals several interesting insights regarding the characteristics of various metrics. While metrics within most coarse categories appear to cluster together in the latent space (e.g., {\tt llite}, {\tt vm}), there are some clear outliers (e.g., {\tt llite\_work\_listxattr}, {\tt vm\_pgmajfault}).

\begin{figure}[htbp]
  \centering
  \captionsetup{justification=raggedright,singlelinecheck=false}
  \includegraphics[width=0.5\textwidth]{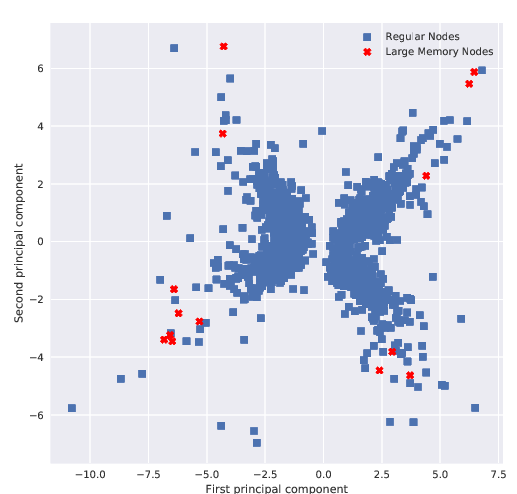}
  \caption{2-dimensional view of the static latent representation for the compute nodes estimated by \sysname\ with $K=10$, using PCA for reduction to two dimensions. Large memory nodes are displayed in a different way. ({\em Best viewed in color.})}
  \label{fig:node_pca}
\end{figure}

In the same way, the static latent representations for the nodes in the matrix ${\bf \bar{U}}$ can be used to visualize the nodes. Figure~\ref{fig:node_pca} shows the two-dimensional plot for the static node behavior. An interesting pattern with two distinct and equal sized clusters is revealed. While a thorough post-analysis of the node specifications is needed to explain the patterns, we distinguish between the regular compute nodes with 24 GB memory and the large memory nodes with 1 TB memory. The large memory nodes are mapped in the periphery of two clusters, indicating that \sysname\ is able to capture the specification related characteristics of the nodes.

\begin{figure}[htbp]
  \centering
  \captionsetup{justification=raggedright,singlelinecheck=false}
  \includegraphics[width=0.5\textwidth]{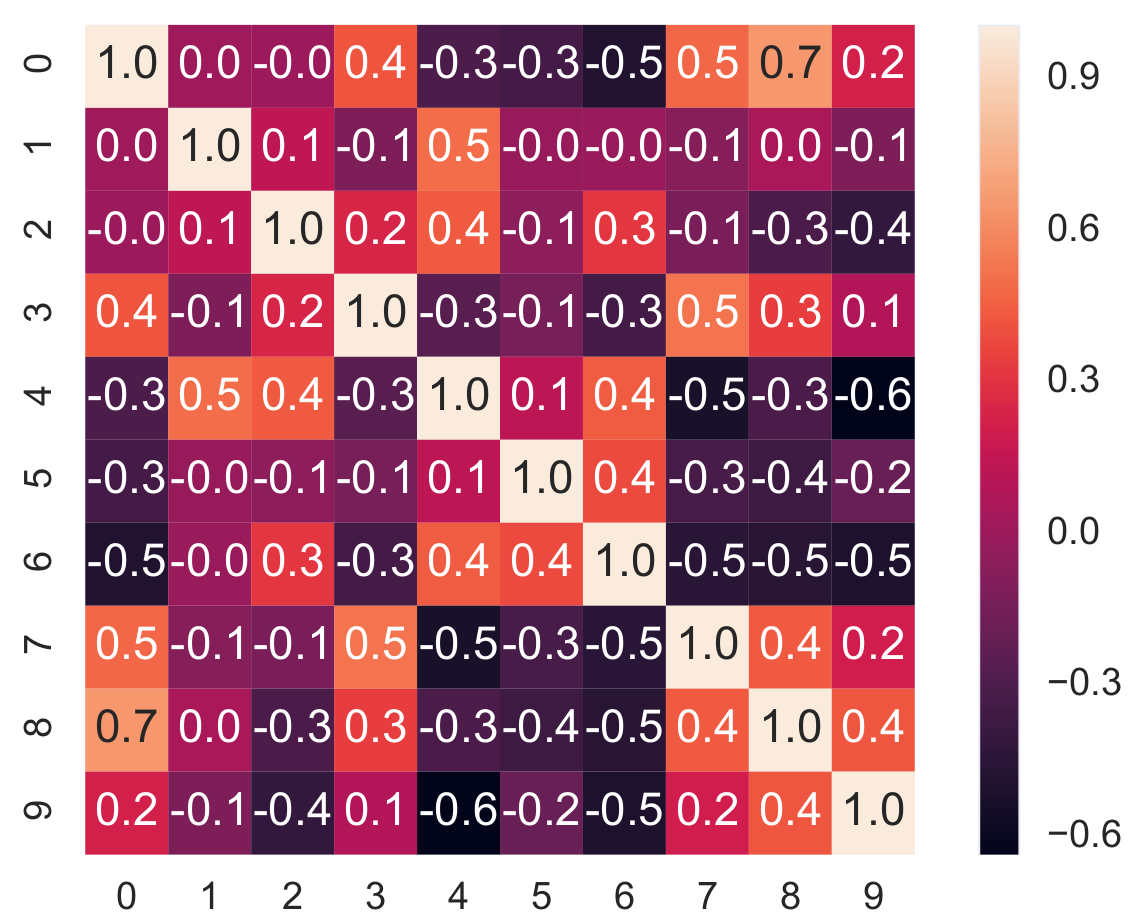}
  \caption{Pearson R correlation among the 10 latent node representations. ({\em Best viewed in color.})}
  \label{fig:node_correlation}
\end{figure}

As mentioned earlier, the key output that is produced by \sysname\, is the dynamic latent representation for each node. An important point here is that each latent dimension contains unique information about the nodes. While we do not explicitly enforce orthogonality in the optimization formulation, the results show that the latent dimensions are not significantly correlated (See Figure~\ref{fig:node_correlation}). 

\begin{figure*}[htbp]
  \centering
  \captionsetup{justification=raggedright,singlelinecheck=false}
  \includegraphics[width=\textwidth]{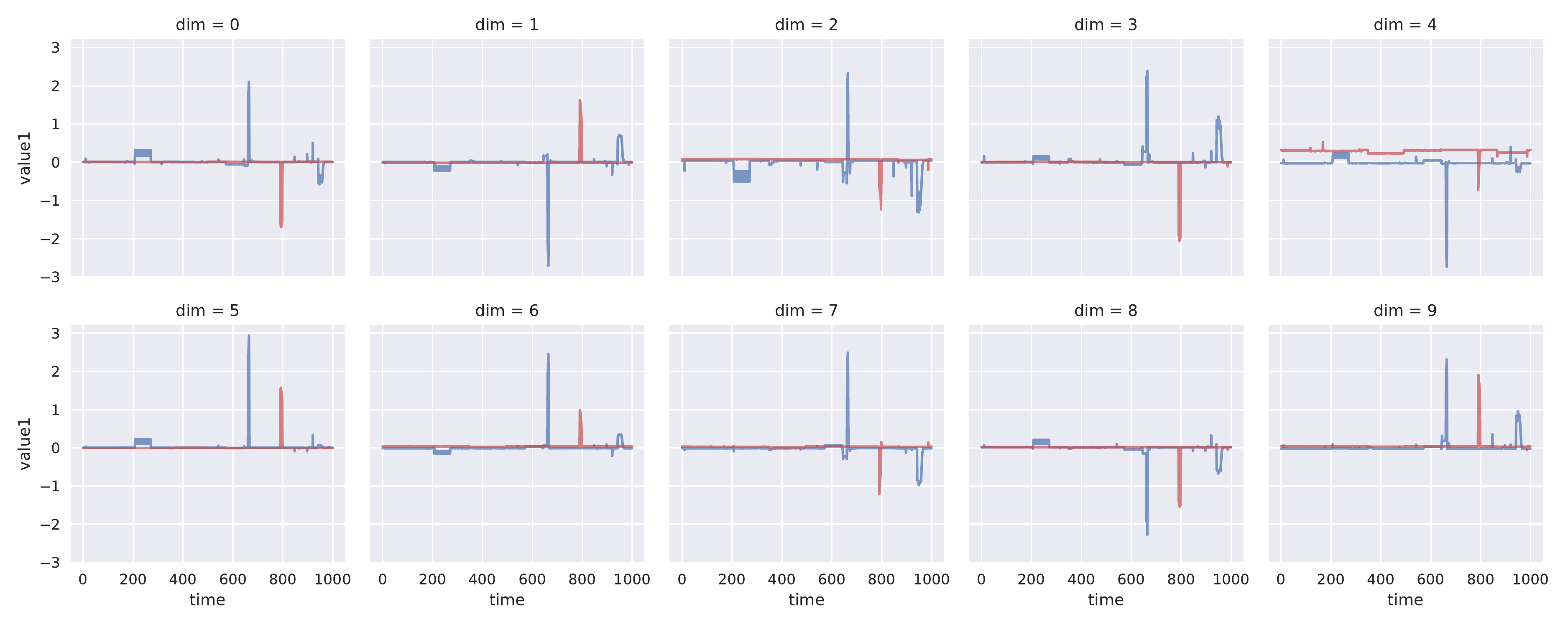}
  \caption{Temporal 10-dimensional representation estimated by \sysname for two selected nodes. Node {\tt c325-312} was inactive during the target week with seven executed jobs, shown in blue plot. Node {\tt c336-203} was the most active during the target week with 60 executed jobs, shown in red plot.. ({\em Best viewed in color.})}
  \label{fig:node_temporal}
\end{figure*}

Figure~\ref{fig:node_temporal} shows the $K$ components of the time-varying ${\widehat{u}_{nt}}$ for two different nodes in the system. For both nodes, the different latent dimensions reveal a unique behavior of the node. Moreover, the output for the two nodes is significantly different. This is due to the fact that the two nodes had significantly different workloads during the target week (obtained from the job information present in the \tacc\ data). Node {\tt c325-312} (blue plot) had a significantly light workload with 9 executed jobs, and consequently the latent dynamic behavior has low variance. On the other hand, node {\tt c336-203} (red plot) had a heavy workload with 60 executed jobs, which is reflected in the high variance for the latent dimensions.
Figure~\ref{fig:node_temporal} shows 10 latent dimensions of the time-varying component ${\widehat{u}_{nt}}$ for two different nodes with two different workloads in the system.
In particular, node {\tt c325-312} (blue plot) had a significantly light workload with seven jobs receiving from users. 
On the other extreme, node {\tt c336-203} (red plot) had the most active jobs during the target week with $60$ jobs in total.
%In Figure~\ref{fig:node_temporal}, these two nodes exhibit similar patterns of temporal behaviors in latent space with sporadic spikes at some certain time steps.
In Figure~\ref{fig:node_temporal}, these two nodes exhibit similar patterns of having sporadic spikes at some certain time steps.
Although it looks counterintuitive to have similar patterns for nodes with two different workloads, one may note that even for node with zero job submitted from users in HPC system, it still has system-level jobs running in the background which make the node have different resource usage at different time, causing sporadic spikes in the temporal behavior as seen in the figure.
In addition, what we observe in Figure~\ref{fig:node_temporal} is the latent temporal representation of nodes after ``explaining away'' the overall time-independent component ${\bf \bar{u}}_n$.
For this reason, spikes appear in the temporal representation as a result of unusual resource consumptions in each node.
%For both nodes, the different latent dimensions reveal a unique behavior of the node. 
%Moreover, the output for the two nodes is significantly different. 
%This is due to the fact that the two nodes had significantly different workloads during the target week (obtained from the job information present in the \tacc\ data). 
%Node {\tt c325-312} (blue plot) had a significantly light workload with 9 executed jobs, and consequently the latent dynamic behavior has low variance. 
%On the other hand, node {\tt c336-203} (green plot) had a heavy workload with 60 executed jobs, which is reflected in the high variance for the latent dimensions.

\subsection{Finding anomalies using \sysname} 
\label{subsec:result3}
In this section, we analyze the association of our resulting hidden values and common errors obtained from system log messages. In particular, there are three frequent error types that we extracted from system log messages: (1) write error, (2) segmentation fault and (3) inode error. Write error is a type of error that happens in the system log with the format ``\textit{ost\_write} operation failed with $-122$". This is a linux error that happens when disk quota is exceeded\footnote{\url{https://github.com/torvalds/linux/blob/master/include/uapi/asm-generic/errno.h}}. Segmentation fault is a common error type when a program tries to write or read in an invalid memory location. Inode error happens when \textit{ll\_inode\_revalidate\_fini()} fails and returns with error code $-43$.
We associate the error types from system log with our resource usage statistics by assigning the error that occurs within 10 minutes of the occurence of the present time window.

In our analysis, we use the anomaly score defined in section~\ref{subsec:anomaly} and analyze its correspondence to the frequent system log error types mentioned above. To compare, we also show results using an adapted anomaly detection method that constructs a three-way tensor using the resource usage data and then uses tensor reconstruction error as the anomaly score for the entire system~\cite{Sorkunlu:2017}. While the original method provides a single score for the whole system, we adapt the method to produce a score for each node individually.

\begin{figure*}[htbp]
	\centering
	\subfloat[write error]{
		\includegraphics[width=0.85\textheight]{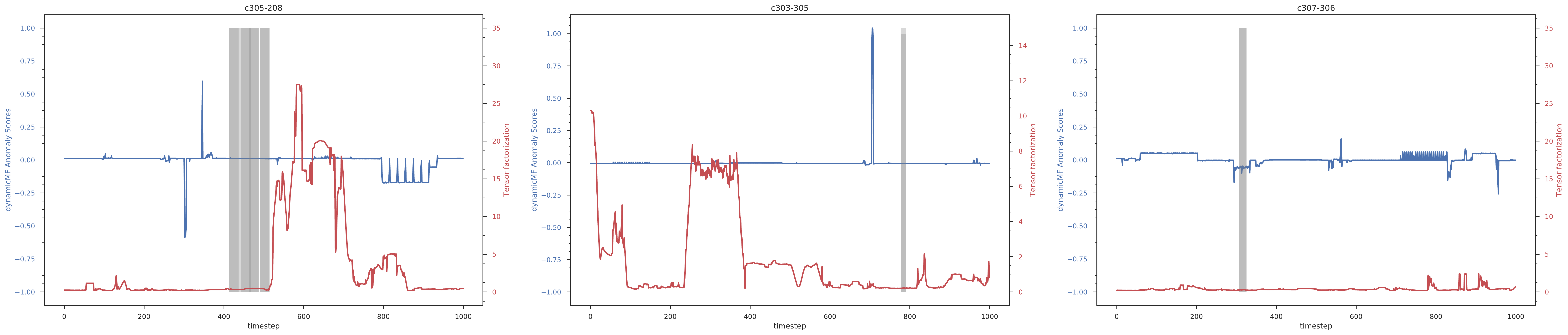}
		\label{fig:write_error}
	}
	\hspace{0mm}
	\subfloat[segmentation fault]{
		\includegraphics[width=0.85\textheight]{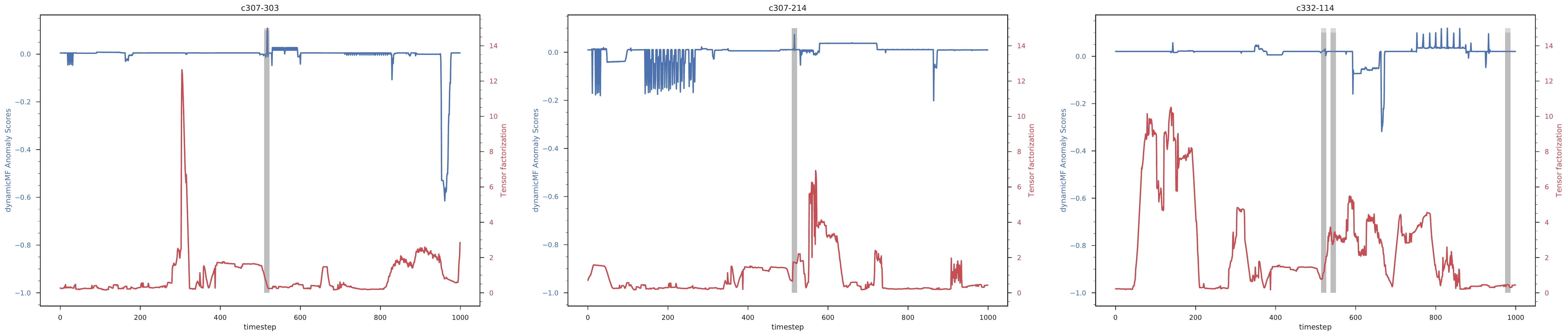}
		\label{fig:segfault_error}
	}
	\hspace{0mm}
	\subfloat[inode error]{
		\includegraphics[width=0.85\textheight]{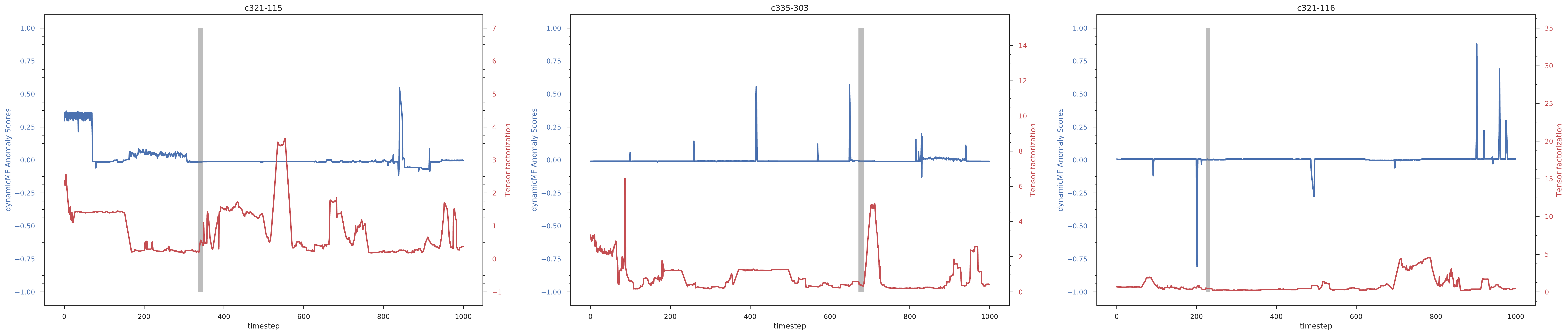}
		\label{fig:inode_error}
	}
	\captionsetup{justification=raggedright,singlelinecheck=false}
	\caption{Association between anomaly scores computed using~\eqref{eq:residual} and errors in system log messages. Anomaly scores produced by \sysname are shown in blue and the anomaly score produced by the state of art tensor reconstruction method is shown in red. Vertical gray lines indicate where the errors happen. The node name is shown above each figure. ({\em Best viewed in color.})}
	\label{fig:error}
\end{figure*}

Figure~\ref{fig:error} shows three typical examples of three computing nodes for each system log error type. In this figure, for write error, we do not observe any association between write error and anomaly score. As the write error happens as a result of exceeding disk quota, this kind of erratic behavior does not reflect in resource consumption and therefore it does not reveal any association with our residual values, as can be seen in Figure~\ref{fig:write_error}. On the other hand, the segmentation faults in Figure~\ref{fig:segfault_error} and inode errors in Figure~\ref{fig:inode_error} occur very close and even identical to the peaks of the anomaly score. As segmentation fault and inode errors are typical examples of buggy programs, these errors are usually results of faulty programming practice. This can lead to the programs with inefficient resource usage, which can be subsequently observed by our resource usage statistics. Therefore, in Figure~\ref{fig:segfault_error} and~\ref{fig:inode_error}, the spikes in anomaly score values often appear roughly around the errors in terms of time.

The tensor based method (red plot in Figure~\ref{fig:error}) produces a metric shows high anomaly values that are temporally close to some of the errors. However, the method also produces a large number of false alarms, which can make such a method practically unusable.

\section{Related Work}
\label{sec:discussion}
The dynamic matrix factorization algorithm proposed here falls under the general purview of {\em unsupervised representation learning}~\cite{Bengio:2012}, that includes a variety of tasks such as dictionary learning, independent component analysis, autoencoders, matrix factorization and various forms of clustering. In particular, a relevant topic is {\em sparse coding}~\cite{Chen:2001}, where the objective is to learn over-complete and sparse bases for a given data set. However, methods in these categories primarily focus on a static representation (typically as a matrix) and do not handle temporally evolving data.

The idea of dynamic matrix factorization has been used in designing the recommendation system. In particular, dynamic Poisson factorization (dPF) model works effectively when we assume each observation follows a Poisson distribution~\cite{charlin2015dynamic}. In this model, because of their probabilistic setting, variational inference has been used to perform inference. After that, they use the block coordinate approach to solve the optimization problem yield as a result of variational inference. Although our model is similar to dPF in the design of shared components and local components, our model is different from dPF in several ways: (1) we do not assume the temporal relationship between consecutive timestep as in dPF, (2) because of our non-probabilistic formulation, we can avoid the problem of non-conjugacy in dPF and use more effective gradient-based optimization such as Adam algorithm instead of using block coordinate optimization with L-BFGS (a more computationally intensive optimization method) as a building block.

Solutions for detecting, diagnosing, and predicting faults and failures in large high performance computing installations have typically relied on message logs~\cite{Xu:2008,Oliner:2008,Fu:2009,Reidemeister:2009,Pelaez:2014} or resource usage data~\cite{Guan:2010,Bronevetsky:2012}, or both~\cite{Chuah:2013:LRU:2553409.2553428,Gurumdimma:2016,Chuah:2016}. Since this paper focuses on the detection task, we present a brief overview of related methods that deal with detecting faults. Methods operating on message logs typically aggregate message logs by an entity of interest, e.g., a computational node/block~\cite{Oliner:2008,Xu:2008} or a job~\cite{Fu:2009}, and identify anomalous entities using different data representations, such as message arrival statistics~\cite{Oliner:2008}, vector representation derived from the message content~\cite{Xu:2008} or a state machine that models the dynamic behavior of the entity~\cite{Fu:2009}. However, message logs are typically noisy and often incomplete, which has led to methods that analyze alternate data sources such as resource usage metrics or performance counter data~\cite{Guan:2010,Bronevetsky:2012}. However, these solutions perform a node-specific or job-specific analysis of resource usage to identify anomalous nodes~\cite{Guan:2010} or jobs~\cite{Bronevetsky:2012}. Recently, solutions that combine resource usage data and message logs to improve fault detection have been proposed~\cite{Gurumdimma:2016,Chuah:2016}. The Crude system uses resource usage data to improve the performance of a PCA driven anomaly detection method~\cite{Xu:2008} that operates at node and job level. However, none of the existing methods model the temporal dimension to better identify faults. In a recent work, a tensor based representation of the resource usage data was used to identify anomalies at the system level~\cite{Sorkunlu:2017}. The method relies on low rank approximation of a three way tensor that captures interactions between nodes, metrics, and time, to establish an anomaly score for the entire system at every time instance. However, the method does not provide a node level estimate of system behavior and anomalies.

\section{Conclusions and Future Directions}
\label{sec:conclusion}
In this study, we propose \sysname\ - a dynamic matrix factorization method which takes both global and local behaviors of the system into account to analyze for variations. This method is applied to a list of time-varying resource usage matrices. In our experiments with real data from Lonestar 4 system, \sysname\ helps reduce the dimension of resource usage statistics and provide visualization at a node-specific level. The proposed anomaly statistic allows for an easy identification of performance anomalies. Future direction of our research will be on developing a fully-automated anomaly detection method to detect failure in HPC system by using the low dimensional latent space obtained from \textit{dynamicMF}.

Future direction of our research will be to leverage job information for analysis and incorporate it into our model. Successfully modeling the interaction between jobs and nodes can help us better explain the data and obtain a more accurate model to detect inefficiencies in resource usage.

\bibliography{references}
\bibliographystyle{abbrv}
\end{document}